\documentclass[runningheads]{cl2emult} 
 
\usepackage{makeidx}  
\usepackage{graphicx} 
\usepackage{subeqnar} 
\usepackage{cropmark} 
\usepackage{eso}      
\makeindex            
 
\newcommand{\gta}{\mathrel{\hbox{\rlap{\lower.55ex \hbox {$\sim$}} 
                   \kern-.3em \raise.4ex \hbox{$>$}}}} 
\newcommand{\lta}{\mathrel{\hbox{\rlap{\lower.55ex \hbox {$\sim$}} 
                   \kern-.3em \raise.4ex \hbox{$<$}}}} 
 
\begin{document} 
\title*{Why Do Black--Hole X-ray Binaries\protect\newline Tend 
to Be Transient?} 
\toctitle{Why Do Black--Hole X-ray Binaries 
\protect\newline Tend to Be Transient?} 
%
%
\titlerunning{Black--Hole X--ray Transients} 
%
\author{Jean-Pierre Lasota} 
\authorrunning{Jean-Pierre Lasota} 
%
%
\institute{Institut d'Astrophysique de Paris\\ 
 98bis Bd Arago, 75014 Paris, France} 
 
\maketitle              
 
\begin{abstract}  

Black hole X--ray binaries are transient probably because their discs
are subject to the same thermal--viscous instability which is present
in dwarf nova binary systems. I discuss applications of the dwarf--nova
instability model to transient, low--mass, X--ray binary systems. When
disc truncation and X--ray irradiation are taken into account this
model is capable of reproducing the basic properties of X--ray binary
outbursts (\cite{dhl}).

\end{abstract} 
 
\section{The thermal--viscous instability} 
All Low Mass X--ray Binaries (LMXBs) in which the presence of a black
hole is inferred from dynamical mass determination are transient. This
could mean that all Black Hole LMXBs (BHLMXBs) are transient, but there
could be non--transient systems  in which black holes have masses in
the range allowed  for neutron stars.  For this reason it is more
prudent to say that BHLMXBs `tend' to be transient (as suggested by the
organizers of this  meeting) rather than to affirm that all of them are
transient.

The responsibility for outbursts in low mass, close binary systems is
commonly imputed (e.g. \cite{mw}) to the usual suspect: the
thermal--viscous instability present in discs of a subclass of
cataclysmic variable stars: the dwarf--nova systems. There, for
temperatures corresponding to hydrogen recombination, changes in
opacity (emissivity) strongly affect cooling mechanism dependence on
temperature and lead to a thermal instability under the standard
assumption about viscous heating. This instability is intimately
related to a viscous instability. A disc's thermal equilibria, at a
given distance from the center, form a characteristic {\sl S} shape in
the surface density -- effective temperature plane. The middle branch
of the {\sl S} corresponds to unstable solutions, the upper one to hot,
almost fully ionized configurations, while the lower branch contains
cold, quasi--neutral states. (The existence of the cold branch does not
depend on the presence of convection, contrary to erroneous assertions
in recent lectures on the subject \cite{ml99}. In fact some attempts to
modify the viscosity prescription required switching off the convection
in order to get a sufficient amplitude of the jump between the cold and
hot branches \cite{ccl}). If the rate at which the mass is transferred
from the black--hole's stellar companion is such that the disc is
(somewhere) unstable, a limit cycle behaviour will follow taking the
system through a cycle of oscillations between hot and cold states.
Such a cycle is supposed to explain dwarf nova and transient LMXB
outburst cycles.

The instability and the limit cycle do not, by themselves, produce
outbursts similar to the ones observed in dwarf novae and transient
systems. The thermal instability creates temperature gradients which
propagate through the disc in the form of heating (transition to the
hot branch) and cooling (transition to the cold branch) fronts. The
form of the resulting luminosity variations depends on what is assumed
about the viscosity or, more precisely, what is assumed about the
viscosity parameter $\alpha$ appearing in the kinematic viscosity
prescription $\nu = \alpha c_s^2/\Omega_K$, where $c_s$ is the speed of
sound and $\Omega_K$ is the Keplerian frequency. Is $\alpha$ is assumed
to be constant, then, because of insufficient temperature and density
contrasts between the cold and hot states, the resulting light--curve
has the form of regular, small amplitude outbursts with no resemblance to
dwarf--nova outbursts or X--ray binary transient events. In order to
obtain dwarf--nova type outbursts the viscosity parameter $\alpha$ must
be assumed to be larger in the hot than in the cold state \cite{s84}.
This {\sl einsatz} has not been, for the moment, based on any model of
viscosity, although a recent argument in favour of different values of
$\alpha$ in hot and cold states \cite{gm} is based on such a model
(beware, however of the erroneous interpretation in \cite{ml99}: it 
applies hot equilibrium disc solutions to describe the properties of a 
cold non--equilibrium disc!).

The assumption of a jump in the $\alpha$ value, however, does not
suffice to reproduce observed properties of outbursts in close binary
systems and additional physical effects have to be added to the `pure'
disc instability model (see e.g. \cite{s99}\cite{hlw}). Some of these
effects, such as inner disc truncation \cite{mnl}\cite{mhln} and disc
X--ray irradiation \cite{dhl}\cite{dlhc}, which are important for low
mass X--ray transients, will be discussed in this article.

Figure {\ref{stab}} shows that the parameters of all BHLMXBs put them
well below the limit for the disc thermal stability. Here I used the
stability criterion for X--ray irradiated discs as proposed by Jan van
Paradijs \cite{vP96}. Irradiation (discussed in Sect. \ref{irr}) of the
outer disc by accretion--produced X--rays from the inner disc gives a
stability criterion which can be expressed in the form (\cite{dlhc},
\cite{l99}):  \begin{eqnarray} \dot M_{\rm transf} > \dot M_{\rm
crit}^{\rm irr}\approx 2.0 \times 10^{15}
		\left({M_1\over M_{\odot}}\right)^{0.5} \left({M_2\over
		M_{\odot}}\right)^{-0.2} P_{\rm hr}^{1.4} \nonumber \\
		\times \left(\frac{\cal C}{5 \times
		10^{-4}}\right)^{-0.5} ~\rm g~s^{-1}
\label{dotp} \end{eqnarray} where $\dot M_{\rm transf}$ is the
mass--transfer rate, $M_1$ is the black hole mass, $M_2$ is the mass of
the companion, $P_{\rm hr}$ the orbital period in hours. ${\cal C}$ is
a parameter describing effects of irradiation (see \cite{dlhc}).
Systems satisfying condition (\ref{dotp}) are stable.

However, systems with $\dot M_{\rm transf} < \dot M_{\rm crit}^{\rm
irr}$ are not necessarily unstable. In fact BHLMXBs are well below the
stability limit (lower than one would expect from evolution models
\cite{mnl}). They are close to the stability limit for cold discs if
their discs are truncated in the inner regions. There are several
arguments in favour of such a truncation.

\section{The necessity of truncation}

According to the model quiescent accretion rate must {\sl everywhere}
satisfy the condition (\cite{hmdlh}):  \begin{equation} \dot M(r) <
{\dot{M}_{\rm cold} = 9.5 ~ 10^{15} ~ \alpha^{0.01} \left( {M_1 \over
\rm M_\odot} \right)^{-0.89} \left( {R \over 10^{10} \; \rm cm}
\right)^{2.68} ~\rm g~s^{-1}} \end{equation}

\begin{figure} 
\centering 
\includegraphics[width=.9\textwidth]{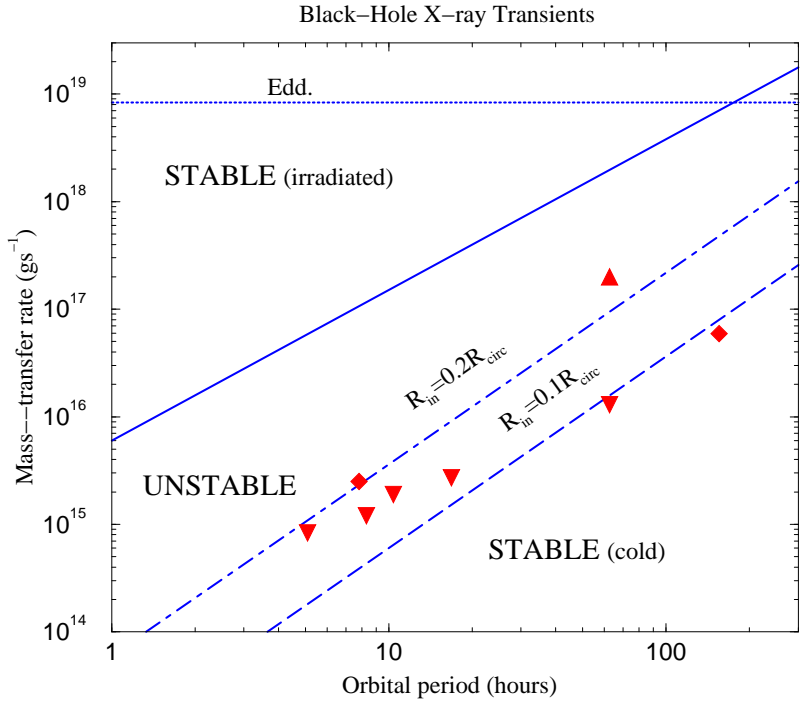} 
\caption[]{Stability criteria for accretion discs in black--hole 
low--mass X--ray binaries. The mass--transfer rate is estimated by dividing 
the mass accreted during outburst by the recurrence time. This mass 
accumulation rate might be slightly lower than the actual transfer 
rate\cite{mnl}. Since, except for two systems, only the lower limits on 
recurrence times are known, we have, for most systems, only the upper limits 
on the mass accumulation rate. All these systems are well below the 
upper line corresponding to stable irradiated discs but are rather close 
to the stability limit for truncated discs, represented here by two 
lines corresponding to truncation radii $R_{\rm in}= 0.1$ and $0.2 
R_{\rm circ}$, where $R_{\rm circ}$ is the circularization radius. Data 
is taken from ref. \cite{mnl}. Two values are given for GRO J1655-40: 
the higher one corresponds to the 1996 outburst (see \cite{elh}). The 
dotted line marked `Edd' corresponds to the Eddington accretion rate 
$\dot M_{\rm Edd}= L_{\rm Edd}/0.1c^2$ for $6 M_{\odot}$.} 
\label{stab} 
\end{figure}

For a disc extending down to the last stable orbit and typical
parameters, this predicts quiescent accretion several orders of
magnitude lower than observed (\cite{l96}). A disc truncated at $10^3 -
10^4 R_S$ (where $R_S$ is the Schwarzschild radius) will solve this
problem, in particular if the inner `hole' is filled by an Advection
Dominated Accretion Flow (ADAF) (e.g. \cite{nbm}, \cite{hlmn},
\cite{emn}). For such truncation radii, what is left of the disc could
be cold and stable as shown by Fig. (\ref{stab}) (where the truncation
radius is scaled by the circularization radius). Also, if we take into
account irradiation, disc truncation is necessary in order to obtain
the light--curves required by observations (\cite{dhl}).

\section{Effects of irradiation} 
\label{irr} 
 
\begin{figure} 
\centering 
\includegraphics[width=.8\textwidth]{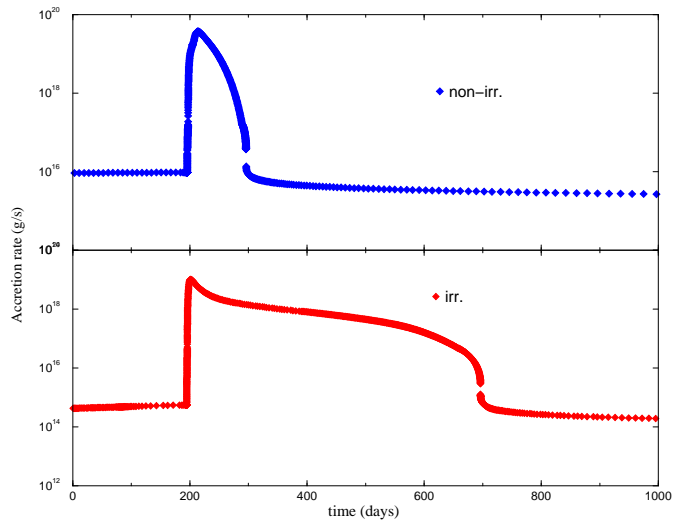} 
\caption[]{Examples of lightcurves obtained in the disc instability 
model for truncated, non--irradiated 
(upper panel) and irradiated (lower panel) discs in black--hole 
low--mass binaries. In the non--irradiated 
case the cooling front rapidly propagates inwards, 
whereas irradiation postpones propagation of this front 
as suggested in \cite{kr}. As a result the light curve 
is close to exponential, outbursts lasts longer and 
more mass is drained from the disc.} 
\label{lc1} 
\end{figure}

There is overwhelming evidence that discs in LMXBs are strongly X--ray 
irradiated (\cite{vPMc}). In outburst, a disc's optical emission is due to 
X-ray reprocessing. As mentioned above, irradiation modifies the 
stability criterion. It also affects all the properties of the outburst 
cycle. Its main effect is to retard the propagation of the cooling front, 
leading to an exponential decay in the light--curves (\cite{kr}). An 
example of this effect is shown in Fig. (\ref{lc1}). A systematic study  
of irradiation effects will be presented in Dubus et al. \cite{dhl}. 
It appears that most of the fundamental properties of at least one 
class of LMXB transient systems can be reproduced by the model if both 
truncation and irradiation are taken into account. The transient systems 
concerned are those with a fast-rise exponential decay (FRED) light curves. 
(See next section for the discussion of recurrence times). 
 
In systems with more complex forms of light--curves one should also take 
into account (the observed) irradiation of the secondary 
(\cite{elh},\cite{e99}). 
 
\section{Recurrence times} 
 
Compared with most dwarf novae, black hole X--ray transients have very 
long recurrence times ($ > 30$ years). These also seem to be longer 
than the recurrence times of 
neutron star transients. Very long recurrence times can be due to 
several effects. First, truncated discs in these systems could be cold and 
stable 
(Fig. (\ref{stab}) and outbursts due to (upward) fluctuations of mass 
transfer. The recurrence time then would correspond to some cycle in 
the secondary star. Such a model was proposed for the dwarf nova WZ Sge, 
whose recurrence time is $\sim 30$ years \cite{lhh}\cite{hlh}. 
 
Second, the inner truncation radius can be such that the 
disc is marginally stable (see Fig. (\ref{stab}). Since for a stable 
disc the recurrence time is 
infinite, for a given mass--transfer rate one can always fine--tune the  
inner disc radius so that the recurrence time is arbitrarily long 
(see e.g. \cite{mhln}. 
This is for example the case of the model proposed in \cite{mhm}. 
Such an assumption is, however, equivalent to assuming global stability 
because the fine tuning would be at the mercy of even very small fluctuations 
of mass transfer. 
 
Two other possibilities can be described by using an approximate formula 
for the recurrence time (see \cite{s93}\cite{mhln}) 
\begin{equation}  
t_{\rm rec} \approx 3   \left(\frac{\xi}{3}\right) \left(\frac{M_1}{M_{\odot}}\right)^{0.62}  
\left( \frac{R}{10^{10}~{\rm cm}}\right)^{0.14} \left( \frac{\alpha_{\rm cold}}{0.02}  
\right)^{-0.83}\left(\frac{T_{\rm eff}}{3000\ {\rm K}}\right)^{-4}  
\ {\rm yr} 
\label{tb}  
\end{equation} 
where $T_{\rm eff}$ is the disc effective temperature, $\alpha_{\rm cold}$ is the  
viscosity parameter in quiescence and $\xi$ is a factor (3 - 5)  
taking into account the fact that the quiescent disc is not in {\sl 
viscous} equilibrium\cite{ilhs}. In a quiescent disc $T_{\rm eff}$ is almost 
independent of the radius (this is {\sl not} the result of low viscosity in 
quiescence, contrary to the erroneous assertions in \cite{ml99}). 
 
We  can see that if we assume $\alpha_{\rm cold}\approx 10^{-3}$, Eq. (\ref{tb} 
gives recurrence times longer than tens of years. This assumes that the 
quiescent effective temperature is not lower than 3000 K. In {\sl non-- 
irradiated} discs this is indeed the case (note that irradiation is important only during the 
outburst) and only by lowering $\alpha_{\rm cold}$ can one obtain 
long recurrence times \cite{mhln}. Irradiation during outburst, however, allows one to 
drain more mass before the cooling front brings the disc into the cold 
state. As a result quiescent temperatures are lower than in the non-- 
irradiated case: $T_{\rm eff} \lta 2000$ K and accordingly recurrence 
times are longer \cite{dhl}.

\section*{Acknowledgments} 
I am grateful to Guillaume Dubus, Anya Esin, Jean-Marie Hameury and Kristen Menou for 
many enlightening discussions. I thank J.-M. Hameury for data used in Fig. 2. 
This work was supported in part by an ASPS/CNRS grant.

\clearpage 
\addcontentsline{toc}{section}{Index} 
\flushbottom 
\printindex 
 

\begin{thebibliography}{7} 
%
\addcontentsline{toc}{section}{References} 
 
 
\bibitem{c98}   Cannizzo, J.K. (1998) ApJ, 494, 318 
 
\bibitem{ccl}   Cannizzo, J.K., Chen, W., Livio, M. (1995), 454, 880 
 
\bibitem{dhl}   Dubus, G., Hameury, J.-M., Lasota, J.-P. (2000) to be submitted to A\&A 
 
\bibitem{dlhc}  Dubus, G., Lasota, J.-P., Hameury, J.-M.,  Charles, P. (1999) MNRAS, 
                303, 139 
 
\bibitem{e99}   Esin, A.A. (2000) these proceedings 

\bibitem{emn}	Esin, A.A., McClintock, J.E., Narayan, R. (1997) ApJ, 489, 865
 
\bibitem{elh}   Esin, A.A., Lasota, J.-P., Hynes, R.I. (2000) A\&A, submitted 
 
\bibitem{gm}    Gammie, C.F., Menou, K. (1998) ApJ, 492, L75 
 
\bibitem{hlh}   Hameury, J.-M., Lasota, J.-P., Hur\'e, J.-M. (1997) MNRAS, 287, 937 
 
\bibitem{hlw}   Hameury, J.-M., Lasota, J.-P., Warner, B. (2000) A\&A, in press 
 
\bibitem{hlmn}  Hameury, J.-M., Lasota, J.-P., McClintock, J.E., 
                Narayan, R. (1997) ApJ, 489, 234 
 
\bibitem{hmdlh} Hameury, J.-M., Menou, K., Dubus, G.,  Lasota, J.-P., Hur\'e, J.-M. (1998) \ \ \ \ \ \  MNRAS, 298, 1048 
 
\bibitem{ilhs}  Idan, I., Lasota,  J.-P., Hameury, J.-M., Shaviv, G., (1999) Phys. Rep. 311, 213  
 
\bibitem{kr}    King, A.R., Ritter, H. (1998) MNRAS, 293, 42 
 
\bibitem{l96}   Lasota, J.-P. (1999) in van Paradijs, J., van den 
                Heuvel, E.P.J, Kuulkers, E. eds., Compact Stars in Binaries, IAU Symp. 
                165, Kluwer, Dordrecht, p. 43 
 
\bibitem{l99}   Lasota, J.-P. (1999) in S. Mineshige, J.C. Wheeler eds. 
                Disk Instabilities in Close Binaries - 25 years of the Disk Instability 
                Model, Universal Academy Press, Tokyo, p. 191 
 
\bibitem{lhh}   Lasota, J.-P., Hameury, J.-M., Hur\'e, J.-M. (1995) A\&A, 302, 29 
 
\bibitem{ml99}  Livio, M. (1999) in J.A. Sellwood, J. Goodman eds., 
                Astrophysical Discs - An EC Summer School, ASP Conference Series Vol. 
                160, p. 33 
 
\bibitem{mnl}   Menou, K., Narayan, R., Lasota, J.-P. (1999) ApJ, 513, 
                811 
 
\bibitem{mhln}  Menou, K., Hameury, J.-M., Lasota, J.-P., Narayan, R. (2000) MNRAS, 
in press 
 
\bibitem{mhm}   Meyer-Hofmeister, E., Meyer, F. (1999) A\&A, 348, 154 
 
\bibitem{mw}    Mineshige, S., Wheeler, J.C. (1989) ApJ, 343, 241 
 
\bibitem{nbm}   Narayan, R., Barret, D., McClintock, J.E. (1997) ApJ, 
                482, 448 
 
\bibitem{s84}   Smak, J. (1984) Acta Astron., 32, 101 
 
\bibitem{s93}   Smak, J. (1993) Acta Astron., 43, 161 
 
\bibitem{s99}   Smak, J. (1999) Acta Astron., 49, 391 
 
\bibitem{vP96}  van Paradijs, J. (1996) ApJ, 464, L139 
 
\bibitem{vPMc}  van Paradijs, J., McClintock, J.E. (1995), in Lewin, 
                W.H.G., van Paradijs, J., van den Heuvel, E.P.J.  eds., X--ray Binaries, 
                CUP, Cambridge, p. 58 
 
 
\end{thebibliography}
\end{document}